\documentclass[twocolumn,epsfig]{revtex4}
\usepackage{epsfig,color}
\begin{document}
\title{Role of asymmetry of the reaction and momentum dependent interactions on the balance energy for neutron rich nuclei}

\author{Aman D. Sood$^1$ }
\email{amandsood@gmail.com}
\address{
$^1$SUBATECH,
Laboratoire de Physique Subatomique et des
Technologies Associ\'ees \\University of Nantes - IN2P3/CNRS - Ecole des Mines
de Nantes 
4 rue Alfred Kastler, F-44072 Nantes, Cedex 03, France}
\date{\today}

\maketitle

\section*{Introduction}
A large number of next-generation radioactive
beam facilities are being constructed or planned in addition to many already existing facilities in the world. At these facilities, nuclear reactions involving nuclei with a large neutron
or proton excess can be studied, thus providing a great opportunity to study both the structure of rare isotopes and the
properties of isospin asymmetric nuclear matter. Complementary to the nuclear structure
studies but being equally important and exciting are reaction studies with radioactive beams, especially heavy-ion reactions induced by neutron-rich beams at intermediate energies.
The ultimate goal of this branch of nuclear physics is to determine the isospin dependence of the in-medium nuclear effective
interactions and the equation of state (EOS) of isospin asymmetric nuclear matter, particularly its isospin-dependent term,
i.e., the density dependence of the nuclear symmetry energy. One has also to look for the ovservables which are sensitive to the symmetry energy since it can not be measured directly from experiments. Collective transverse in-plane flow as well as its disappearance has been found to be the one of the most sensitive observables to dynamics of heavy-ion collisions at intermediate energies \cite{gautam2}. In recent study Gautam et al., \cite{gautam7} has found that in-plane flow is sensitive to symmetry energy as well as its density dependence. Moreover the sensitivity to symmetry energy increases with increase in neutron content of the system. Here we present a systematic study of disappearance of flow i.e. balance energy $E_{bal}$ for an isotopic series of Ca with N/Z varying from 1 to 2 for different density dependences of symmetry energies. We also extend this study for asymmetric reactions having radioactive projectile and stable target. The present study is carried out using IQMD model \cite{hart98}.

\section*{Results and Discussion}
We have simulated several thousand events at incident energies around
 balance energy in small steps of 10 MeV/nucleon for each
isotopic system of Ca+Ca having N/Z (N/A) varying from 1.0 to 2.0 (0.5-0.67).
To check the sensitivity of N/Z (N/A) dependence of E$_{bal}$ towards
density dependence of symmetry energy, we have calculated the
E$_{bal}$ throughout the isotopic series for different forms of
symmetry energy The various forms are F$_{1}(u) \propto u$, F$_{2}(u) \propto u^{0.4}$, F$_{3}(u) \propto u^{2}$ where\emph{ u} = $\frac{\rho}{\rho_{0}}$
In fig. 1a (b) we display the N/Z (N/A) dependence of
E$_{bal}$ fir symmetric reactions of Ca isotops with N/Z (N/A) varying from 1-2 (0.5-0.67) for different forms of symmetry energy; F$_{1}$(u)
(solid circles), F$_{2}$(u) (diamonds), and F$_{3}$(u)
(pentagons). For all the cases E$_{bal}$ follows a linear
behavior. Clearly, N/Z (N/A) dependence of E$_{bal}$ is sensitive to the
density dependence of symmetry energy. We also note that for a fixed N/Z (N/A) stiff symmetry energy F$_{1}$(u) shows
less E$_{bal}$ as compared to soft F$_{2}$(u) whereas super stiff
symmetry energy F$_{3}$(u) shows more E$_{bal}$ as compared to
F$_{2}$(u). For detailed discussion on this we refer to \cite{sood4}
\begin{figure}[!t] \centering
\vskip 0.5cm
\includegraphics[angle=0,width=6cm]{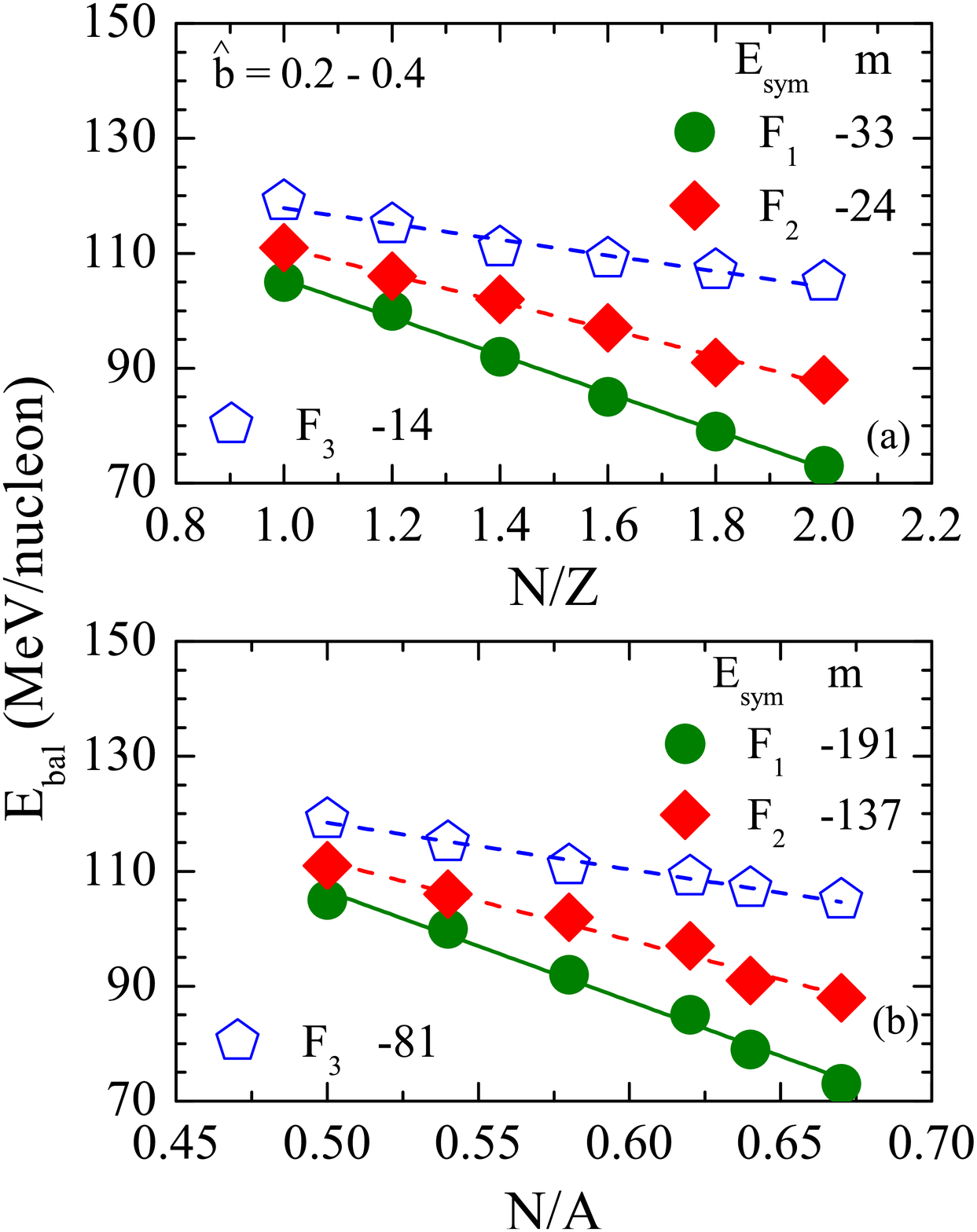}
\caption{\label{fig1} E$_{bal}$ as a function of N/Z (upper panel) and N/A (lower panel) of system
Various lines and symbols are explained in the text.}
\end{figure}
Since one cannot use radioactive isotopes as targets,
therefore, as a next step we fix the target as a stable isotope
$^{40}$Ca and vary the projectile from $^{40}$Ca  to $^{60}$Ca and
calculate E$_{bal}$. In this case the N/Z (N/A) of the reaction varies
between 1 to 1.5 (0.5 to 0.6) and the asymmetry $\delta = \frac {A_{1}-A_{2}}{A_{1}+A_{2}}$
of the reaction varies from 0
to 0.2. The results are displayed by solid green stars in figs. 2(a) and (b) (upper panels).
The solid green circles represent the calculations for symmetric
reactions. Lines represent the linear fit $\propto$ m. 
\begin{figure}[!t] \centering
\vskip 0.5cm
\includegraphics[angle=0,width=6cm]{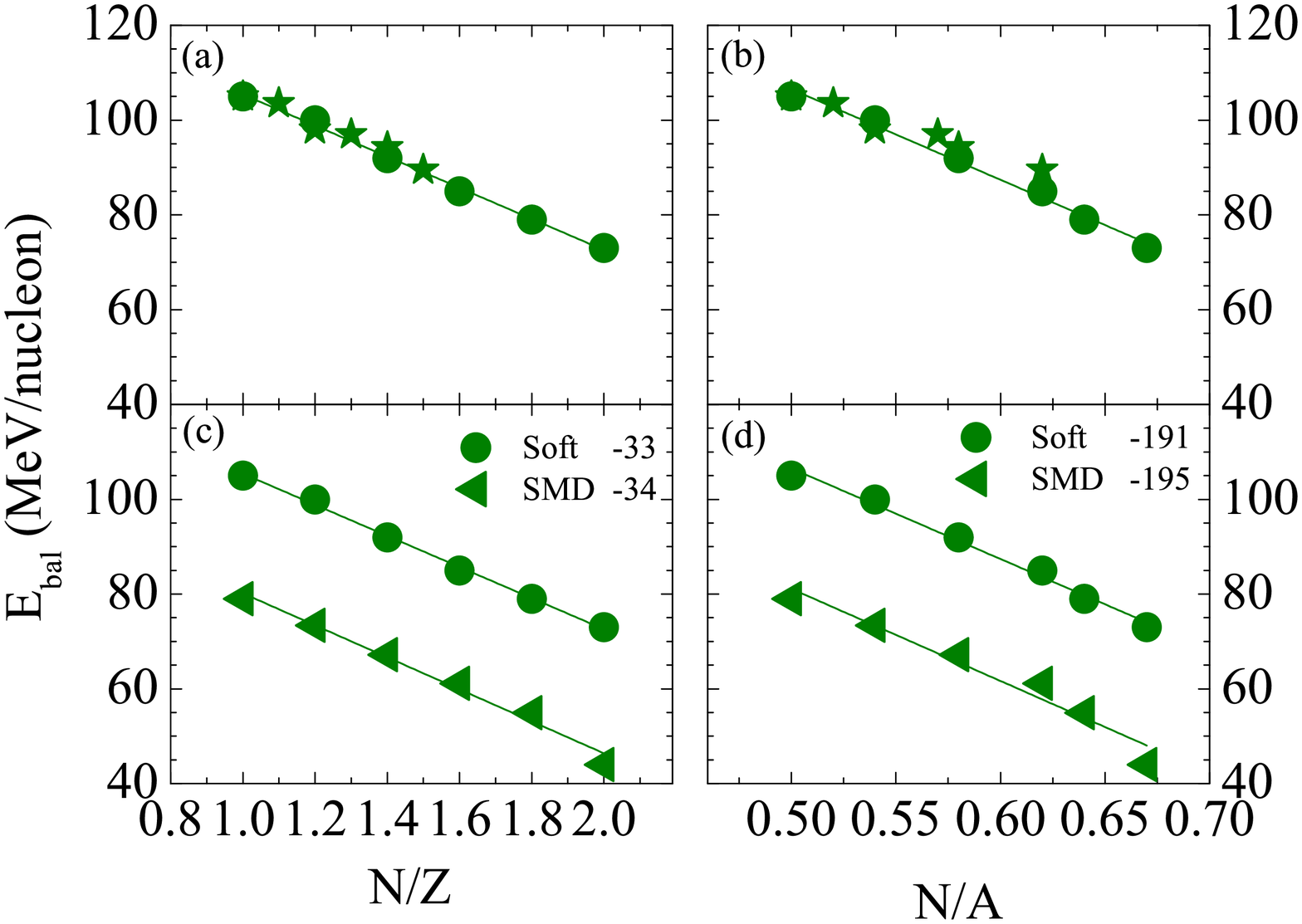}
\caption{\label{fig1} E$_{bal}$ as a function of N/Z (left panel) and N/A (right panel) of system. Various lines and symbols are explained in the text.}
\end{figure}
We see that N/Z (N/A) dependence of E$_{bal}$ is same for both the
cases. We also find that as
the N/Z (N/A) decreases from 2 (0.67) in case of $^{60}$Ca+$^{60}$Ca  to 1.5 (0.6) for
$^{60}$Ca+$^{40}$Ca, the E$_{bal}$ also decreases accordingly. Now the
E$_{bal}$ for
$^{60}$Ca+$^{40}$Ca has same value as in case of symmetric reactions with
N/Z (N/A) = 1.5 (0.6) i.e. the value of E$_{bal}$ is decided by the N/Z (N/A) of the system and is independent of the asymmetry
of the reaction.
It has also been reported in literature that the momentum dependent interaction (MDI) affects
drastically the collective flow as well as its disappearance \cite{soodmdi}. To
check the influence of MDI on the N/Z (N/A) dependence of E$_{bal}$ we
calculate the E$_{bal}$ for the whole N/Z (N/A) range from 1 to 2 (0.5 to 0.67) for the symmetric reactions with
SMD equation of state and symmetry potential F$_{1}(u)$. The
results are shown in figs. 2(c) and (d) (lower panels) by solid left triangles. We find that although the MDI
changes drastically the absolute value of E$_{bal}$ (by about
30\%), however the N/Z (N/A) dependence of  E$_{bal}$ remains unchanged
on inclusion of MDI. Therefore, the dependence of  E$_{bal}$ as a
function of N/Z (N/A) on the symmetry energies of other different forms
(F$_{2}(u)$ and F$_{3}(u)$) is also expected to be preserved on
inclusion of MDI.
\section*{Acknowledgments}
This work has been supported by a grant from Indo-French Centre for the Promotion of Advanced Research (IFCPAR) under project no 4104-1.


\end{document}